# Subwavelength Meta-Waveguide Filters and Meta-Ports

M. Khatibi Moghaddam[1], Romain Fleury[1,*]

[1] Laboratory of Wave Engineering, Ecole Polytechnique Fédérale de Lausanne (EPFL), Lausanne, 1015 Lausanne, Switzerland

This paper proposes a novel technique for the design of miniaturized waveguide filters based on locally resonant metamaterials (LRMs). We implement ultra-small metamaterial filters (Meta-filters) by exploiting a subwavelength (sub-$\lambda$) guiding mechanism in evanescent hollow waveguides, which are loaded by small resonators. In particular, we use composite pin-pipe waveguides (CPPWs) built from a hollow metallic pipe loaded by a set of resonant pins, which are spaced by deep subwavelength distances. We demonstrate that in such structures, multiple resonant scattering nucleates a sub-$\lambda$ mode with a customizable bandwidth below the induced hybridization bandgap (HBG) of the LRM. The sub-$\lambda$ guided mode and the HBG, respectively, induce pass- and rejection- bands in a finite-length CPPW, creating a filter whose main properties are largely decoupled from the specific arrangement of the resonant inclusions. To guarantee compatibility with existing technologies, we propose a unique subwavelength method to match the small CPPW filters to standard waveguide interfaces, which we call a meta-port. Finally, we build and test a family of low- and high-order ultra-compact aluminum CPPW filters in the Ku-band (10-18GHz). Our measurements demonstrate the customizability of the bandwidth and the robustness of the passband against geometrical scaling. The 3D-printed prototypes, which are one order of magnitude smaller and lighter than traditional filters and are also compatible with standard waveguide interfaces, may find applications in future satellite systems and 5G infrastructures.

## I. INTRODUCTION

Over the past few years, the development of new generations of satellite systems, such as nano/micro/cube satellites, has fueled a considerable search for miniaturized microwave/mm-wave systems. Microwave filters are the main building blocks of passive devices, which typically rely on microstrip, dielectric, coaxial, and waveguide technologies. Waveguide filters are the ideal solution for space applications due to low energy loss and high-power handling capabilities, with no frequency limitation [1]. The main drawback of waveguide filters is their large volume and high weight due to the size of their constitutive waveguide cavities, which directly scale with the operating wavelength ($\lambda$). These cavities are coupled using E-plane or H-plane irises, stubs, or posts [2–4]. Among various efforts to make microwave filters smaller, the coaxial combline filters [4]; and evanescent waveguide filters [5] have established themselves as compact microwave filters for space applications, respectively, with coaxial and waveguide interfaces. Combline filters have reduced dimensions because they use parallel-coupled coaxial cavities, where cascaded coaxial cavities implement a chain of TEM modes coupled with self and mutual capacitances.

On the other hand, evanescent filters rely on the evanescent modes of a hollow waveguide, below cut-off [5], loaded by an array of ridges, spaced in $\lambda/4$ distances. Both filters are implemented in a hollow metallic waveguide or box, whose width is smaller than the width of a standard waveguide ($\lambda/2$). The operating frequency of these filters strongly depends on the air gap between the ridges/rods and the top plate [5], and their overall transverse size and length are strongly tied to the operating frequency.

---

*romain.fleury.epfl.ch

This work aims to demonstrate a new exciting opportunity to design and manufacture ultra-small waveguide metamaterial filters (Meta-filters), whose small footprints are largely decoupled from the operating frequency, by using the strong non-local interactions enabled within metamaterials. Metamaterials are artificial wave media, structured at subwavelength scales, where the collective action of the constitutive elements (meta-atoms) triggers effective properties not readily found among natural materials [6]. Metamaterials are typically studied using homogenization approaches, where the mesoscopic properties of the composite are exploited for the creation of various devices in a wide range of frequencies [7–10]. Locally resonant metamaterials (LRMs), in particular, are interesting since they employ local resonant inclusions to create a medium with strong frequency and spatial dispersion. LRMs consist of a subwavelength ensemble of small local resonators embedded in a propagating host, which enables strong multiple scattering down to the deep subwavelength scale in a wide variety of non-planar platforms, from microwaves to acoustics [11]. For instance, the interaction of the continuum plane waves and local resonances makes an anti-crossing, which induces a hybridization bandgap (HBG) around the local resonance frequency of the constitutive elements. The HBG can be exploited for blocking microwaves over ultra-small volumes or manipulating them over small electrical scales by resonant doping [12–14]. They have enabled ultra-small cavities and different types of waveguides, which localize modes inside the HBG, from acoustic to optical frequencies [13,15–17]. However, these solutions all have a narrow bandwidth and high group velocity dispersion, and cannot be used for waveguide filter design.

Here, we propose and experimentally demonstrate a new approach for using LRMs to provide a guided mode, and accordingly, a meta-filter with customizable bandwidth, in an ultra-small footprint that does not scale with the wavelength, while being compatible with standard interfaces used in current waveguide technology. For this purpose, instead of building an LRM with free space as a host, we load a hollow metallic waveguide or pipe with local resonators and leverage the sub-$\lambda$ mode of such LRM waveguide (LRMW). Remarkably, the sub-$\lambda$ guided mode, which falls below the HBG of the loaded LRM, has an adjustable bandwidth, enabling us to build custom bandpass filters. The LRMW can be implemented by arranging small uniaxial metallic elements, such as wires or pins, helical, spiral, or ring resonators spaced at deep subwavelength distances. To demonstrate the features of LRMWs, in most sections of the paper, we use an exemplary model built by inserting ultra-thin pins inside a hollow rectangular pipe, named a composite pin-pipe waveguide (CPPW). In all sections of the paper, we target the Ku-band (10-18GHz), commonly used for satellite communications. We demonstrate that the size of the pins can be modified for tuning a specific operating frequency, while the width of the host pipe adjusts the bandwidth independently. We also show that the induced HBG provides a sharp frequency selection and high level of rejection in the upper side of the passband, where the positions of the passband and rejection bands are decoupled from the arrangement of the resonant pins and geometrical scaling. We design and simulate three coaxial CPPWs with various types of hollow pipes and resonators. Finally, we investigate a solution to make an adapter between the miniaturized CPPW filter and the standard waveguide interfaces, and propose a specific type of CPPW port, named metamaterial port (meta-port), which can be used to improve the matching in a subwavelength volume. These results are confirmed with experiments on low and high-order CPPW filters, which are fabricated by selective laser melting using an aluminum alloy. Our findings demonstrate the customizability, compactness, and ideal RF metrics of CPPWs, allowing these miniaturized components to be used in a wide range of frequencies and bandwidths for space and terrestrial applications.

## II. RESULTS

### A. Dispersion engineering of LRMWs

As shown in [13,15,16], in volumetric LRMs, built by arranging electrically small resonators with sub-$\lambda$ separations in a host medium, the interaction between the continuum of plane waves and local resonators allows efficient wave manipulation over sub-$\lambda$ distances (Fig. 1(a)). The possibility to block wave transmission or impart large phase delays over electrically small distances can be intuitively understood near the resonance frequency of the local resonators since the large phase variations of the scattered field near the resonance result in Fano interferences that give birth to a high-index frequency band followed by an HBG [12,13,17]. The nucleation of the HBG can also be pictured as a classical equivalent of quantum polaritons [14,18,19], an anti-crossing between a continuum of states and an individual resonance. In LRMs, the lower and upper polariton-like modes are obtained, respectively, below and above the HBG, due to multiple resonant scattering coming from the resonant meta-atoms. Very different from Bragg band gaps, the position of the HBG is relatively independent of the periodicity of the medium and is dominantly determined by the resonant behavior of the inclusions. However, the width of the HBG depends on the strength of multiple scattering and the density of inclusions.

At microwave frequencies, the wire medium, which is often described as a low-frequency plasma [20,21], has been employed to create an HBG, and guide electromagnetic signals along defect lines made from additional shorter wires, resonating in the HBG[13]. However, the slow guided-mode associated with such a line-defect is highly dispersive, and its bandwidth cannot substantially be manipulated. The technique proposed here for realizing custom waveguide filters can also be based on wires, pins, or other small local resonators (Fig. 1(a)), but the crucial difference is that they are inserted inside a dispersive host: a hollow metallic waveguide, schematically represented in Fig. 1(b). As we shall see, the parameters of the dispersive host, namely its cut-off frequency $f_c$, characteristic impedance $Z_0$, and propagation constant $k$, bring crucial degrees of freedom to design practical filters with customizable properties. The obtained LRM waveguide (LRMW), shown in Fig. 1(c), is very different from LRMs in free space in terms of its ability to create a guided mode. In this paper, we will mostly focus on the archetypal case displayed in Fig. 1(d), namely a one-dimensional (1D) periodic LRMW, in which the local inclusions can be modeled by identical admittances $Y_r=iB_r$ forming periodic parallel loads inside the host transmission line. We note $f_r$ the resonance frequency of the inclusions and $a$ the period, with $a\ll\lambda$. We use a simple analytical model to show that, generally, resonant periodic inclusions inside a host waveguide can lead to two different propagation regimes. For simplifying this initial discussion, we assume that resonators are only couple via the modes of the host (direct coupling between the resonators will be taken into account later when a more advanced model is needed). This LRMW, accordingly, has a constant propagation $\gamma=\alpha+j\beta$ which depends on $Z_0$, $k$, $a$, and $B_r$, as derived in [22] for any periodic medium:

$$\cosh(\gamma a) = \cos(ka) - \frac{1}{2} B_r . Z_0 \sin(ka). \qquad (1)$$

This relation allows us to identify the two operating regimes that can occur if we operate near the resonance frequency of the inclusions. Whenever $|\cos(ka) - \frac{1}{2} B_r . Z \sinh(ka)| \geq 1$, $\gamma$ is purely real, and the structure only supports attenuating waves. Written at resonance ($B_r=0$), this relation leads to $|\cos(ka)| \geq 1$, which implies that $k$ is imaginary. Thus, the unloaded waveguide must support propagating waves at that frequency for this attenuating regime of the LRMW to be possible: loading a propagative waveguide with resonant inclusions induces a bandgap near their resonance frequency. On the other hand, the opposite case of a propagative LRMW implies that $\gamma$ should be imaginary, and this happens when $|\cos(ka)| \leq 1$, namely that $k$ is purely real. Thus, the host waveguide must be operated in the evanescent regime: below cut-off, resonant inclusions can open a passband into an otherwise attenuating waveguide.

With this in mind, we expect LRMWs to come into two classes: depending on whether the host is propagative or evanescent at the resonance frequency of the inclusions. Fig. 1(e) and (f) show the band structures in the cases of propagative and evanescent host pipes, respectively. Considering the first regime, if the unloaded host waveguide supports propagative waves around $f_r$, i.e., $f_r > f_c$ (Fig. 1(e)), we obtain the nucleation of an HBG near $f_r$, surrounded by lower and upper polaritons (LP, UP). The LP mode, which goes from $f_c$ up to the lower edge of the HBG, is a sub-$\lambda$ mode since its dispersion curve is below the one of the host waveguide: its phase varies over smaller distances than in the host, especially near the upper band edge. In the second regime, we load an evanescent host medium whose cut-off frequency ($f_c$) is larger than $f_r$ (Fig. 1(f)). The result is the creation of a sub-$\lambda$ guided mode, whose bandwidth $BW$ depends on how close the resonance is from the cut-off frequency. Fig. 1(f) represents two cases on the same plot, showing that adjusting $f_c$ controls the bandwidth of the sub-$\lambda$ mode. By decreasing the cut-off frequency of the unloaded host waveguide from $f_{c2}$ to $f_{c1}$, we can extend the lower edge of the band, implying a larger bandwidth ($BW_1 > BW_2$). Another interesting feature is the creation of a HBG above this band due to hybridization with the evanescent modes, which boosts the natural evanescent behavior of the host in the HBG frequency range. This suggests the use of this second regime for small filters with customizable bandwidth and sharp roll-off near the band edges. In the next section, we demonstrate these hypotheses using a 1D LRMW, composed of an array of pins inside a hollow pipe, named composite pin-pipe waveguides (CPPW), in the frequency range of 10-18GHz.

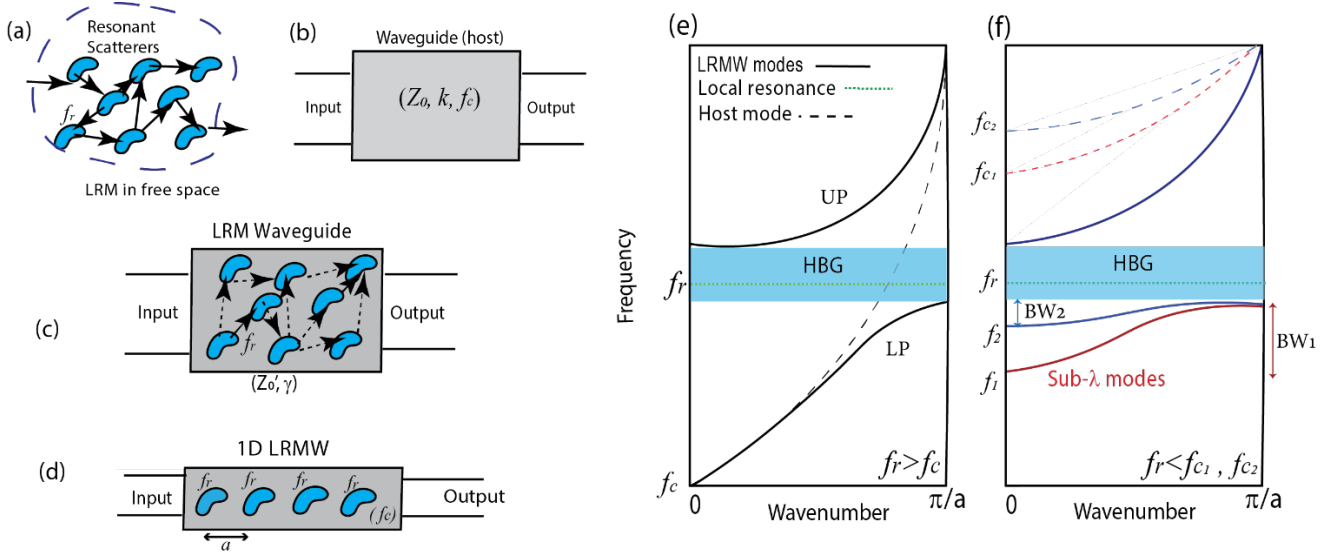

Fig. 1 Concept of locally-resonant metamaterial waveguides (LRMW). (a) A cluster of local resonators in free space induce strong multiple scattering near their resonance frequency $f_r$; (b) A single-mode host waveguide with characteristic impedance $Z_0$, propagation constant $k$, and cut-off frequency $f_c$; (c) LRMW created by loading an ensemble of subwavelength resonators inside a host waveguide; (d) A simpler version of LRMW using a 1D periodic array of resonance inclusions, with a subwavelength periodicity of $a$. (e,f) The schematic of the band diagrams of 1D LRMWs in two different regimes: (e) the host waveguide is above cut-off ($f_r > f_c$) and (f) the host waveguide is evanescent ($f_r < f_c$). In the second case, two different cut-off frequencies are considered, where $f_{c1} > f_{c2}$, leading to different bandwidths $BW_2 < BW_1$.

## B. Characterization of the CPPWs

We now implement a subwavelength LRMW in a realistic structure, namely a composite pin-pipe waveguide (CPPW), composed of a host rectangular metallic waveguide loaded with a deep subwavelength arrangement of pins of height $h_r$, radius $r$, and periodicity $a$, where $r \ll a$ and $a \ll \lambda$. The pins touch only the bottom wall of the waveguide and resonate when $h_r$ approaches the quarter-wavelength condition. To adjust the cut-off frequency $f_c=c/2W$, we play with the waveguide width $W$ keeping a fixed waveguide height of $h$=9.52mm, which corresponds to the height of the standard WR75 waveguide. The WR75 standard is commonly used for satellite communication systems in the X- and Ku-band (10-18GHz). We set $h_r$=5mm, so that the pins resonate in the target frequency range ($f_r$=15GHz), and assume a pin diameter $2r$ between 0.5mm and 1mm, which is close to the minimum thickness that can be fabricated by selecting laser melting (SLM) using a low-loss aluminum alloy of AlSi10Mg, according to current technological standards.

Let us now check the occurrence of the two above-mentioned regimes: *i*) $2h_r<W$, the host waveguide supports propagative TE$_{10}$ mode around $f_r$; *ii*) $2h_r>W$, the host waveguide is evanescent around $f_r$. A unit cell of an infinite system with a length $a$=2.5mm, with periodic boundary conditions (PBCs) along *x*, is shown in Fig. 2(a). The metal is assumed to be a perfect electrical conductor (PEC). For the first regime, we choose $W$=19mm, which is the standard WR75 width, where $f_r$=15GHz falls above the cut-off $f_c$=7.9GHz. The band structure, obtained by finite-element simulations, is plotted in Fig. 2(b) with a dashed black line. Consistent with our expectations, an HBG is created, and a sub-λ mode is observed starting from $f_c$=7.9GHz to the lower edge of the HBG, close to $f_r$. In the second (evanescent) regime, keeping the pin sizes to $h_r$=5mm, we use a narrower width $W$=7mm ($f_c$=21.5GHz) to meet the condition $2h_r>W_2$. As shown by the red dispersion bands in Fig. 2(b), a sub-λ mode is induced below $f_c$, from 11.5GHz to 13.7GHz. The electric field of this mode (LP$_1$), shown in Fig. 2(c), is localized to the pins with a field distribution that allows the energy to be transmitted from pin to pin. We therefore expect this mode to create a passband below the HBG. On the other hand, the UP$_2$ mode, as shown in Fig. 2(c), is not localized, and it is perpendicular to the incident TE modes of the waveguide ports. Thus, the UP$_2$ mode cannot be efficiently excited from the ports and will not create a passband above the HBG.

After this modal study, we design a finite-length CPPW by connecting an array of six unit-cells to standard WR75 waveguide ports, as shown in Fig. 2(d). This component, which acts as a 6$^{th}$ order filter, is made from six pins with $h_r$=5mm separated by distances of $a$=2.5mm. We perform a full-wave simulation to extract the transmission coefficient $S_{21}$ of the two-port system in the two considered regimes. The transmission spectra of Fig. 2(e) are in complete agreement with the eigenvalue studies. The S$_{21}$ of the device for $W$=19 shows the perfect transmission of the TE$_{10}$ mode above 8GHz and a deep stopband in the HBG region. On the other hand, the CPPW in the second regime ($W$=7mm) supports a narrower passband, which results from the sub-λ mode below the HBG (LP) and has a sharper roll-off at 14GHz. It also confirms that the UP band above the HBG does not create a good passband; accordingly, the rejection band expands in a wide range of frequencies from 14 to 20 GHz. This CPPW structure ($W$=7mm), with passband in the frequency range of 10-13GHz, has a total length of 13.5mm, implying an overall size of 0.45$\lambda$ ($\lambda$=30mm), which is roughly an order of magnitude smaller than commercially available filters in this frequency range [23]. Assuming two waveguide ports (with an arbitrary length of 9mm), at both sides, the overall length of the component is only 31.5mm, which is 35% of the size of the most compact Ku-band waveguide filter that we found on the market [24]. To obtain a reasonable impedance matching in the passband, the first and the last pins of the filter are inserted at the boundary between the CPPW and the adapters. Matching can be improved further by designing compact meta-ports, as we will demonstrate in section D.

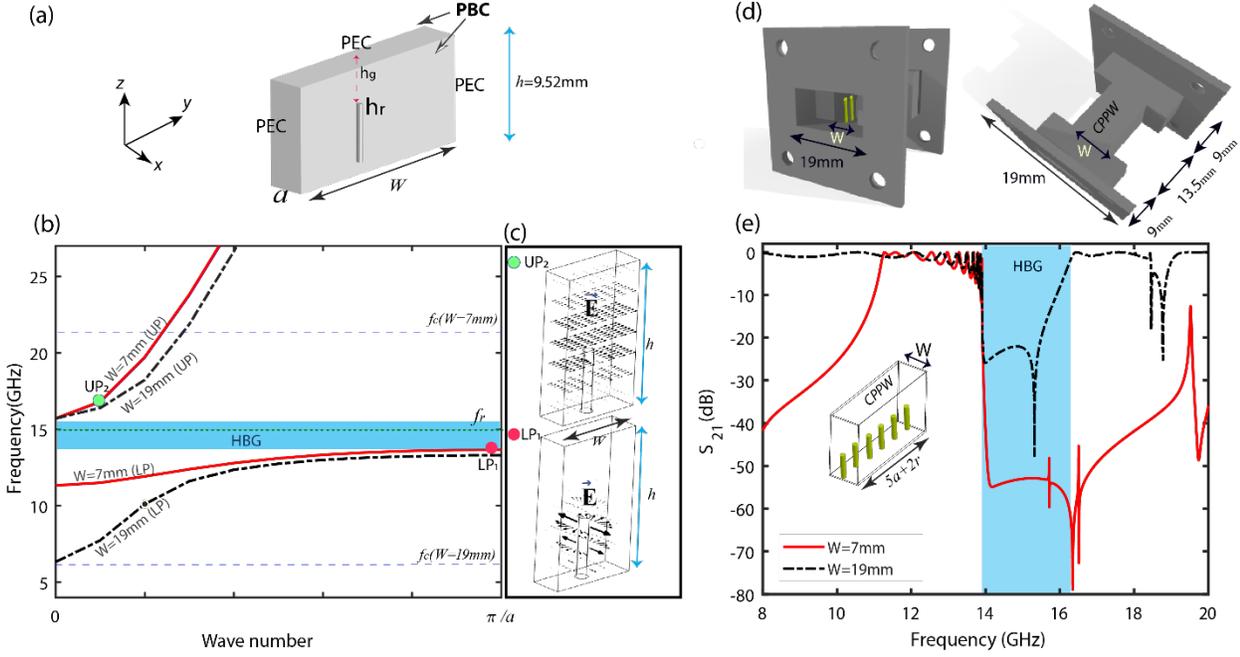

Fig. 2. Implementation of a waveguide-based LRMW. (a) A unit cell of a 1D CCPW structure is made by one ultra-thin pin, with height $h_r$, inserted in a rectangular waveguide of length a<<λ with a width of W. The pin and the bottom wall, made of a perfect electrical conductor (PEC), are in contact, and periodic boundary conditions (PBCs) are along x. (b) The corresponding band diagram for $W$=19mm (propagative host) and W=7mm (evanescent host). (c) The electric field distributions of the lower and upper (LP and HP) modes for the evanescent regime (W=7mm). (d) A CPPW waveguide filter is formed using six unit cells directly connected to standard WR75 waveguide adapters (19 mm width), on both sides. (e) The transmission spectra of the filter for the two different regimes. The hybridization bandgap (HBG) is shown by the shaded blue region.

From now on, we exclusively focus on the second regime, based on an evanescent host. In Appendix I, we detail our analytical modeling and highlight the guiding mechanism of the sub-λ mode in this regime, which allows us to calculate the evanescent coupling coefficient ($\kappa_{evan}$) between adjacent pins as a function of the main geometrical parameters. The formula shows that the coupling $\kappa_{evan}$ strongly depends on the width of the pipe $W$, and thus $W$ is the parameter having the most impact on the guided-mode bandwidth. To illustrate the effect of $W$, we compute the band diagram for various values of this parameter, shown in Fig. 3(a), assuming $a$=2.5mm, $h_r$=5mm, and $h_g$=4.52mm ($h$=9.52mm). It shows that the lower edge of the HBG ($f_o$) does not depend on $W$, contrary to the bandwidth of the guided mode (passband). Besides, as shown in Fig. 3(b), the position of the passband can be tuned by altering the heights of pins. Consistent with the predictions of our analytical model (Appendix II), varying $a$ slightly affects the coupling coefficient. Larger values of $a$ lead to relatively narrower bandwidths, as confirmed by the full-wave simulations of Fig. 3(c). Finally, Fig. 3(d) shows only a slight shift of the guided mode band when altering pins radii, where $r$ varies from 0.1mm to 0.5mm. These behaviors are very different from what is typically observed in traditional evanescent and combline filters, which are sensitive to the cross-section of rods and ridges. A deeper comparison between our method and combline filters is given in Appendix II.

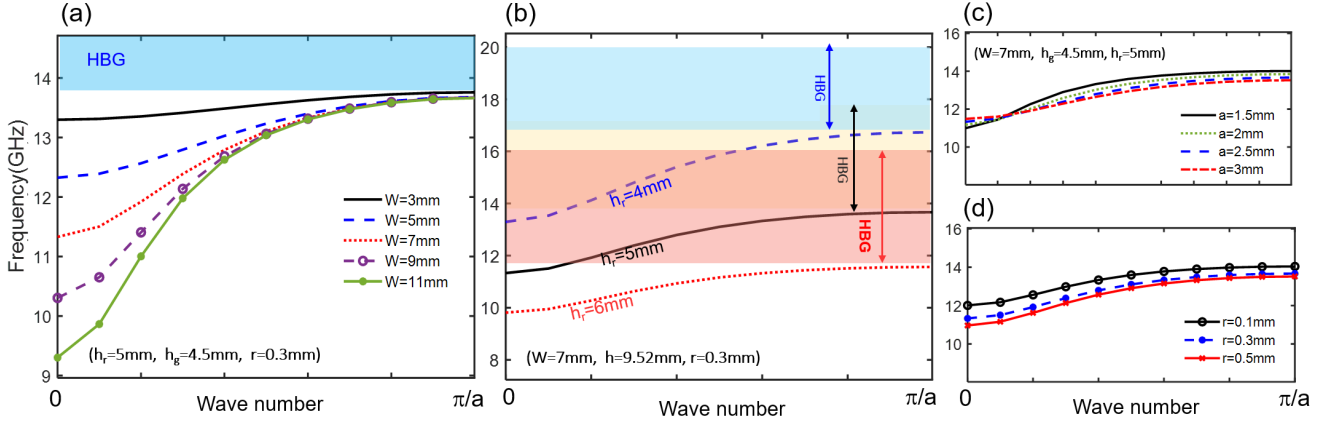

Fig. 3 Variation of the guided mode dispersion band for various values of: (a), the width of the pipe $W$; (b), the height of the pins $h_r$; (c), the period $a$; and (d), the radius of the pins $r$. The shaded region shows the HBG. For each panel, the values of fixed parameters are indicated inside parentheses.

We now confirm that the main tuning knobs of the band structure have similar direct effects on the transmission spectra of finite length filters. First, we compute the scattering parameters ($S_{21}$, $S_{11}$) of a filter built by loading a single meta-atom (pin) in a narrow-width pipe, where $h_r$=5mm. This filter is the smallest instance of a single-pole CPPW bandpass filter, and it is shown in Fig. 4(a). This 1st order CPPW filter is examined for various values of $a$ (2-3.5mm), and $W$ (3 and 7mm). The transmission scattering parameter $S_{21}$ confirms that for smaller values of $W$, this single-pole filter has sharper roll-off and higher rejection level, while the bandwidth is reduced.

Figure 4(a) also exhibits larger bandwidth and lower insertion loss (IL= 20log10 ($S_{21}$)) in the passband, for smaller values of $a$. This information helps us to design high-order CPPW filters, where we need to consider a tradeoff between the size $a$ and the selectivity of the filter. Thus, we can minimize the insertion loss (IL(dB)= -$S_{21}$(dB)) by reducing $a$, but at the cost of decreasing the sharpness of the filter and the rejection level.

By stacking multiple identical cells, each additional pin adds a zero to the upper rejection band, which can be interpreted as the gradual formation of the HBG. By increasing the number of pins to more than five, a sharp roll-off between the guided mode and the HBG is obtained at a converged frequency, $f_o$=13.7GHz, and increasing further the number of pins does not change this frequency. Besides, $f_o$ is also robust against varying $W$. As demonstrated in Fig. 4(b), for a 6th order CPPW filter ($h_r$=5mm, $a$=2.5mm), increasing $W$ results in enlarging the bandwidth of the CPPW filter without moving the location of the HBG. Our studies indicate that by setting the upper cut-off frequency around 14GHz, altering $W$ from 3mm to 16mm results in a fractional bandwidth variation from 3% to 80%, which is a truly remarkable property for a device with a constant, subwavelength footprint.

We now demonstrate that the position and bandwidth of the passband are largely decoupled from the exact arrangement of the pins or a geometrical scaling of the component along to the direction of the wave path. For this purpose, we now consider a CPPW filter with one row of randomly positioned pins (Fig. 5(a)), with maximum deviations of Δy=±2mm, Δx=±1mm from the periodic 1D array. For the two other filters, we put two rows of periodic and randomly positioned pins (Fig. 5(b) and (c)), respectively comprising two rows of pins with fixed distances $a_x$=2.5mm or two rows of pins arranged with a maximum random deviation of Δy=±2mm, Δx=±1mm. Finally, we simulate a squeezed 1D periodic CPPW, for which the entire geometry is scaled-down by 20% in the guiding direction (Fig. 5(d)).

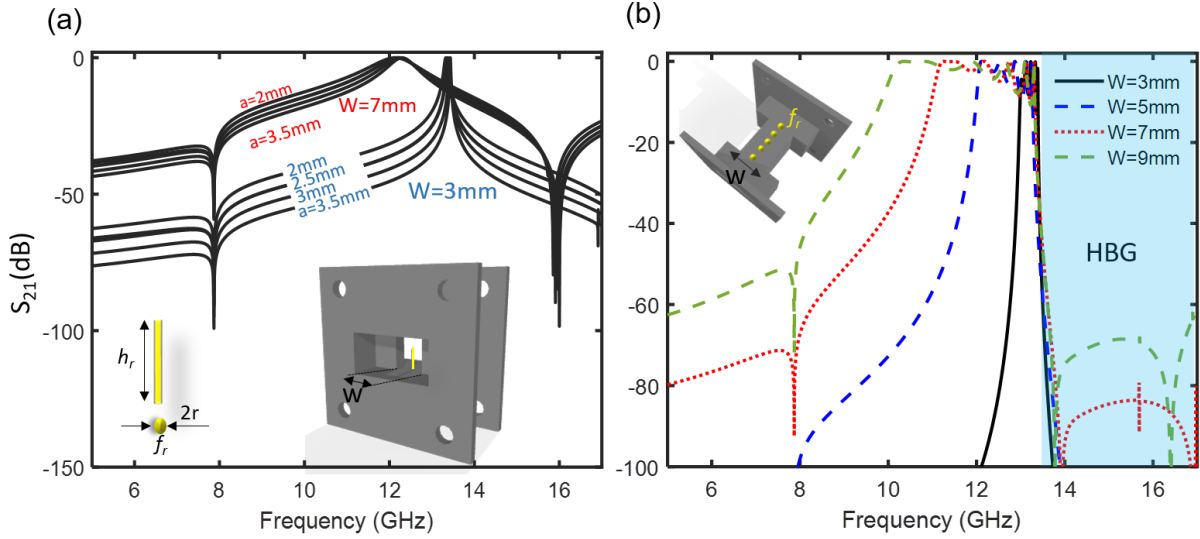

Fig. 4. Bandwidth tuneability. The figure shows the transmission spectra of (a) a single pin CPPW filter for various values of $W$ and $a$, and (b) the 6th order CPPW filter for various values of $W$, with sharp selectivity at $f_o$.

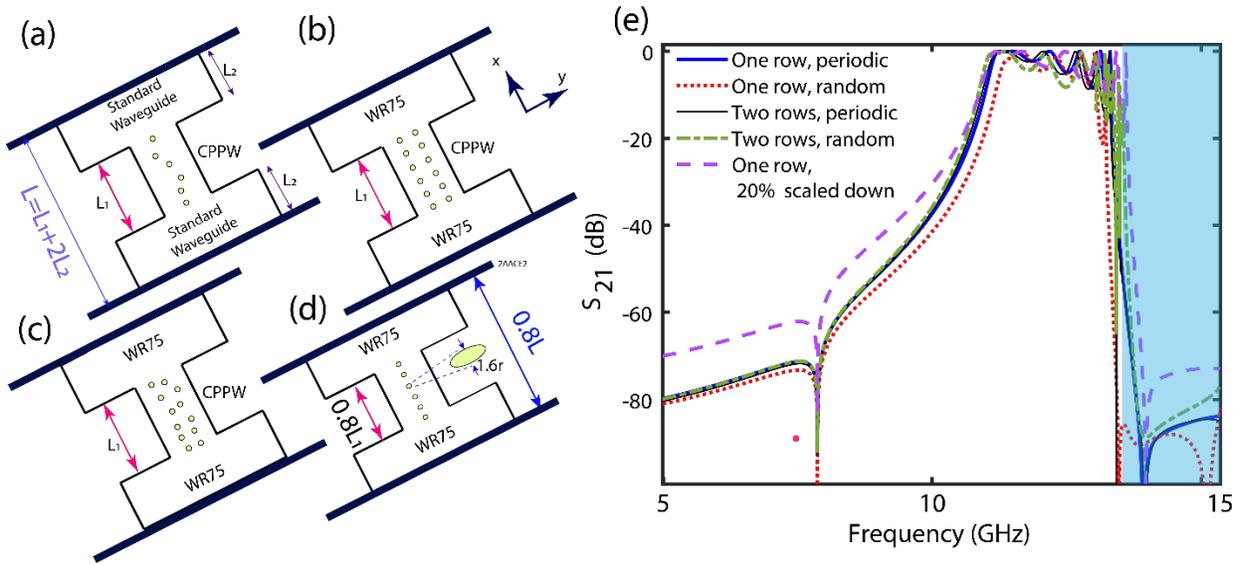

Fig. 5. CPPW filters with standard waveguide ports (WR75), considering $h_r$=5mm, $a$=2.5mm. (a) One row of pins, which are randomly positioned around y-axis with maximum deviation $\Delta y=\pm 2$mm and $\Delta x=\pm 1$mm from a periodic 1D array; (b) 2D array of periodic pins, comprising two rows of pins with distances $a_x$=2.5mm; (c) two rows of pins, which are arranged with a maximum random deviation of $\Delta y=\pm 2$mm, and $\Delta x=\pm 1$mm; (d) a squeezed CPPW for which all geometrical parameters are scaled-down by 80% in the guiding direction (y-axis), making the device shorter and the pin cross-section oval. (e) The scattering transmission spectra ($S_{21}$) of a for all these various pin arrangements.

In all of these cases, we assume $W$=7mm, $r$= 0.3mm, and $h_r$=5mm. The $S_{21}$ of these structures are shown in Fig. 5(e), where all instances still support a passband at the same position with negligible differences in passband ripples and relatively small shifts around $f_o$. These findings exhibit the very stark contrast between CPPW filters and traditional filters, which are typically extremely sensitive to the arrangement of the elements inserted to create the poles and zeros of the transfer function.

## C. Other design possibilities

By design, CPPWs are also compatible with coaxial ports and various types of other hollow metallic hosts, with different cross-sectional shapes, where the bandwidth of each CPPW is determined merely by the relation between the cut-off frequency of the host pipe and the resonance frequency of the inclusions. In coaxial filters, we use surface mount SMA connectors, extending their inner conducting line through two holes in the top metallic wall of the CPPWs, located at a distance of 2.5mm from the first and last pins. Different types of host pipes can support a HBG in the same frequency range, as long as the same pin size is used. For example, rectangular and circular CPPWs, shown in Fig. 6(a) and (b), respectively with diameters of 7mm and 10mm, have equal pass- and stop-bands. Besides, in order to form a host waveguide, we can use an artificial bandgap material instead of metallic walls, such as the artificial PMC walls used for gap waveguides (GWs) [25,26]. As a side note, notice that GWs are very different from the technique proposed here: while they use metamaterials for realizing artificial PMC walls, their waveguide modes and the resonant cavities that are used are similar to those found in conventional waveguide technology, and thus they are not suited to manipulate waves within deep-subwavelength volumes. To illustrate a case of CPPW based on artificial walls, we build a 6$^{th}$ order CPPW filter (Fig. 6(c)), where 2D arrays of rods, which are 7mm apart, are inserted around the line of resonant pins. These rods, with the size of $h_w$=8mm, creates a large HBG above 9GHz. By design, the passband of the CPPW filter falls inside the bandgap of these artificial walls. The transmission spectrum, shown in Fig. 6(d) together with the one of the other cases discussed in this section, exhibits a large HBG above 13.8GHz. However, the rejection band of this filter cannot go beyond the upper frequency of the artificial walls bandgap, which ends at 16GHz. Although the rejection band of this filter is not as wide as the one of a hollow pipe, it has a sharper roll-off before the passband due to resonances within the artificial walls, which effectively increase the filter order. The position and size of the artificial walls bandgap can be modified by altering the density and length of the rods ($h_w$).

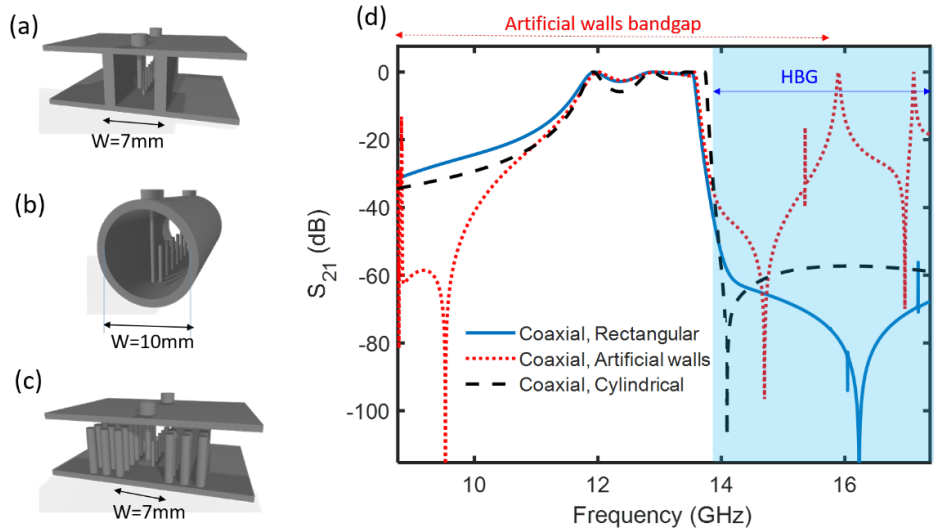

Fig. 6. Different host waveguides. Example of 6$^{th}$ order CPPW filter with coaxial ports, based on (a) rectangular metallic pipe, compatible with planar structures, (b) cylindrical metallic pipe with a diameter of 10mm, (c) bandgap material walls, where the walls have rods with length $h_w$ =7mm. (d) The transmission scattering spectra ($S_{21}$) of all these different CPPWs.

The capability of LRMs for creating bandpass filters in the evanescent regime is not limited to resonators in the form of wires or pins. Figure 7 indeed reports similar transmission spectra obtained with helical resonators and rings. We adjusted the size of these small electrical resonators such that their resonance frequencies both occur near 14GHz.

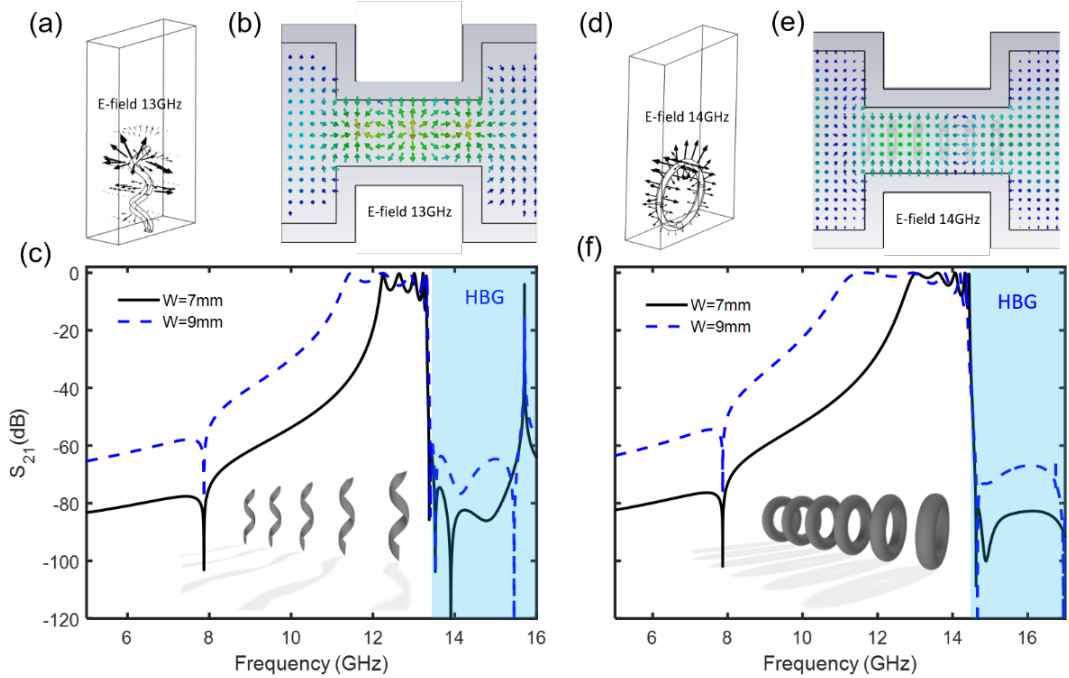

Fig. 7. Different resonators. Transmission spectrum of alternative LRMWs compatible with WR75 standard waveguide using: (a) parallel uniaxial 1D periodic helical wires with two turns, a height of 4.2mm and diameter of 1.5mm. (b) typical field distribution in the passband, (c) transmission spectra for two different values for the pipe width. (d-f) Same as (a-c) but for 1D periodic parallel rings (torus), with diameters of 5mm.

### D. Subwavelength metamaterial waveguide ports

To create a miniaturized CPPW filter compatible with standard waveguide systems, we must provide an improved matching between a narrow CPPW and a standard waveguide port. Here, we consider a WR75 (19.05mm×9.52mm) standard waveguide interface connected to a 6$^{th}$ order CPPW. Since the mode profile of a CPPW waveguide is much narrower than the standard waveguide, the direct connection of WR75 to the CPPW induces a considerable reflection and results in low matching efficiency, which lead to passband ripples in the $S_{21}$ spectra of Figs. 2(e), 4(b) and 5(e). Among different methods that we have investigated for improving the wave transition and reduced insertion loss, the following technique has proven itself to be particularly relevant, as it is realized in a subwavelength volume.

We construct a short length CPPW inside the WR75 transition, by loading it with pins resonating above the WR75 cut-off to make a novel metamaterial port, which we refer to as meta-port. Our goal is to increase the S21 and smooth the ripples seen in Fig. 2(e) near the upper edge of the pass band, around 13-14 GHz. As explained in the first section, if we insert a single pin (with size $h_p$) inside a WR75 waveguide, where $2h_p<W'(W'=19.05mm)$, it induces a drop point (zero) in the transmission spectrum. We choose the size of $h_p$ slightly smaller than the size of the pins of the CPPW filter designed in previous sections ($h_p <h_r$), so that it resonates beyond the edge of the passband. Such a structure, shown in Fig. 8, can be assumed as a notch waveguide filter, which makes a zero at $f_p$. Clearly, the association of this notch filter with the CPPW filter will add a zero to the upper rejection band, without changing its functionality, probably even improving its roll-off. However, because it modifies the field near the transition, it creates an opportunity to improve the impedance matching.

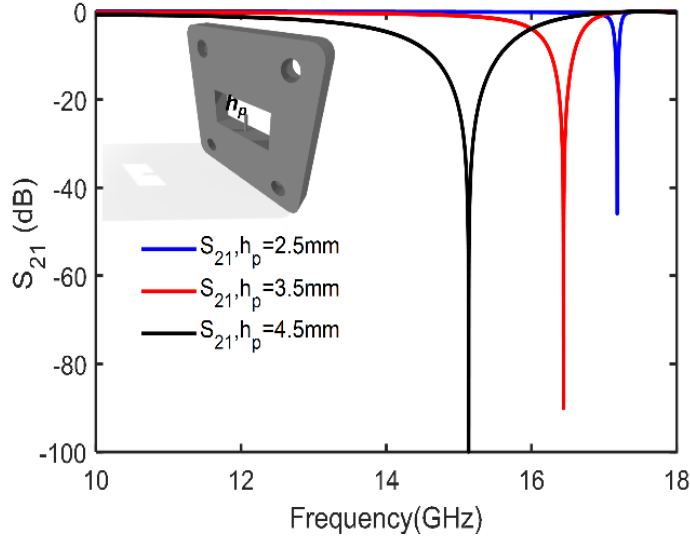

Fig. 8. A short length WR75 waveguide, loaded by a resonant pin with size $h_p$, where $h_p < h_r$. The transmission spectrum, $S_{21}$, for various values of $h_p$.

For this purpose, we do not use a single resonant pin but a cluster of them, and we place it inside the WR75 port, with the goal of shaping the field distribution near the transition between the WR75 and the CPPW sections. Figure 8(a) shows a square array of four pins (2×2) with height $h_p$=3.5mm and separating distances $a_p$=2.5mm, placed in a standard WR75 (we do not yet connect it to the CPPW filter). The transmission spectrum ($S_{21}$) of this structure (label A) shows a stopband around 18 GHz. However, this waveguide efficiently transmits around 13-14GHz. The polarization of the pins in this frequency range (see inset for the energy distribution near 13.7GHz) is not the same as the modes near the band edge (around 18GHz). The field distribution is also different from the one of the typical $TE_{10}$ WR75 modes, since it has subwavelength variations and more concentration around the pins. By connecting this array of four pins to an evanescent waveguide with a similar width as our CPPW filter ($W$=7mm, $L_e$=2.5mm), labeled as structure B, the transmission is decreased in the whole frequency range since the second waveguide does not support propagation below 20GHz (Fig. 9(a)). Next, we put one pin with height $h_r$=5mm in the evanescent waveguide, near the transition, which resonates at the ripple frequency (around 13.7 GHz). This pin, placed in an evanescent host (the transition) makes a 1st order CPPW filter (label C), with a transmission pole around 13.7GHz (Fig. 9(b)). As seen in the inset, the meta-port squeezes the electric field distribution to match the WR75 mode and the narrower CPPW. Obviously, there is a tradeoff between the upper roll-off and the insertion loss that has to be considered when choosing an ideal value for $h_p$. In our case, meta-ports with $h_p$=3.5mm result in a minimum passband insertion loss.

Now, we use such transitions on both sides of a CPPW, with $h_r$=5.15mm, to make a bandpass filter for 11-13GHz (Fig. 10(a)). As expected, these two meta-ports work as adapters to realize a field distribution that can match the mode profile of the CPPW filter (Fig. 10(b)) and accordingly induce ripple-free transmission in the passband, while improving the level of rejection in the upper rejection band. This is evident in the full-wave simulation results of Fig. 10(c) and (d), showing the $S_{11}$ and $S_{21}$ coefficients of the CPPW filter, with and without WR75 ports, respectively.

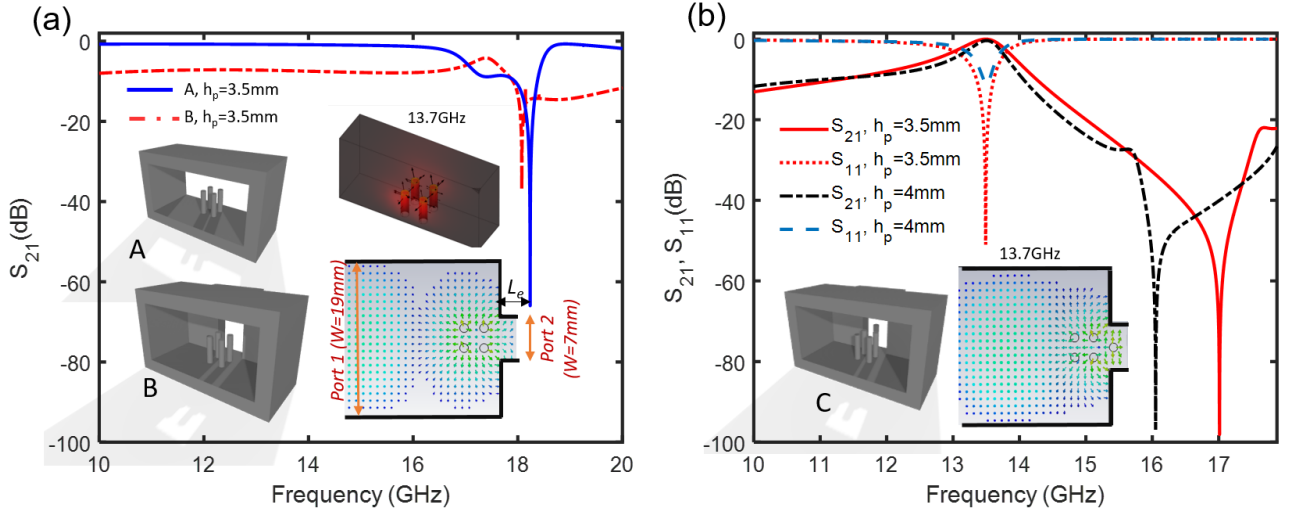

Fig. 9. Designing meta-ports. The figure shows transmission spectra for structures built by inserting 2×2 arrays of pins with height $h_p$=3.5mm (a) inside and near one standard port WR75 (A) and in the proximity of the connection of WR75 to an evanescent waveguide with $W$=7mm (B); (b) adjacent to the connection border of WR75 and a one-pin CPPW filter (C).

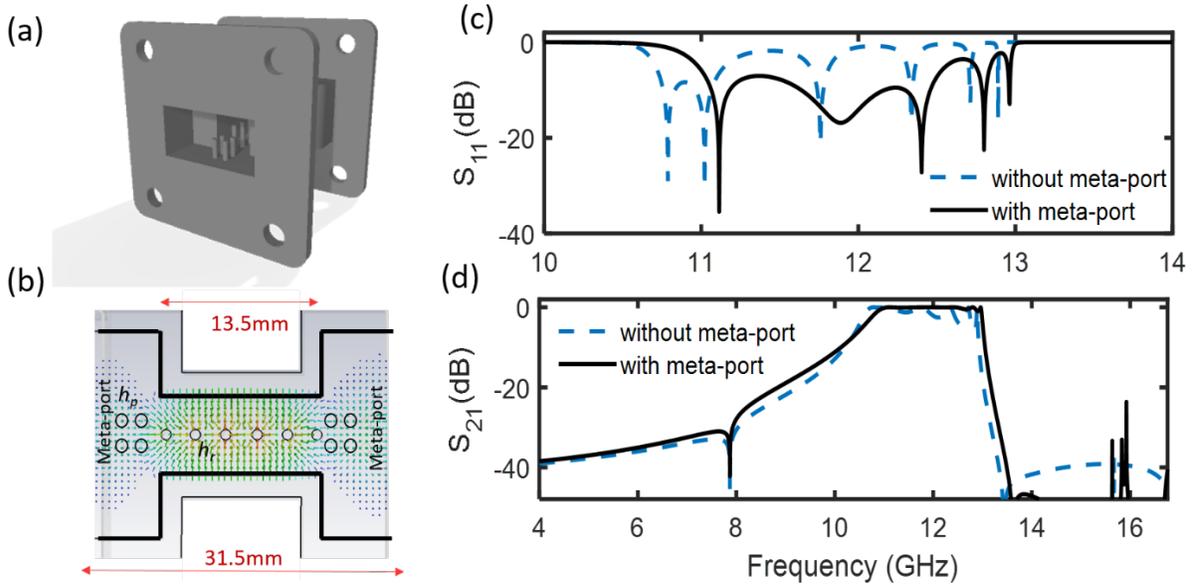

Fig. 10. The efficiency of the designed meta-port in a realistic filter. (a) A 6$^{th}$ order CPPW filter with $h_r$=5.15mm, connected to meta-ports at both sides having $h_p$=3.5mm, (b) the field distribution of CPPW filter and meta-ports and CPPW filter at 12.5 GHz. The effect of meta-port on (c) $S_{11}$ and (d) $S_{21}$ of the CPPW filter.

### E. Fabrication and experiment

We initially designed and fabricated a two-pin 2$^{nd}$-order Ku band CPPW filter at 13GHz (Fig. 11(a) and (b)), directly connected to a standard WR75 waveguide with square flanges (no meta-port). Among different manufacturing techniques for such a metallic component, we chose the selective laser melting (SLM) process with AlSi10Mg aluminum alloy because of its good conductivity, relatively high fabrication speed, and low weight [27]. This Aluminum metal 3D printing technique supports fine details as small as 0.5mm, a minimum wall thickness of 1mm, a dimensional tolerance of ±0.2mm, and matt and glossy finishing. Fig. 11(a) and (b) show pictures of the prototype, in which the length of the primary host pipe is 2.5mm (~λ/10), and the width is 3mm, aiming at a 2% bandwidth. The total size of the component, comprising two pins, is 9mm (0.39λ), and the weight is 7.4g, considering 1mm for the diameter of pins. Fig. 11(c) shows the results of the characterization of the prototype, that agree well with measurements.

To create a wideband filter without changing the length of CPPW, we increase the width, choosing $W$=5mm and 7mm. We also add meta-ports on both sides to improve the matching and use thicker walls to avoid dissipations caused by wavy shapes on thin flanges. The whole length of the two samples, which have the same footprint, is 12.5mm, shown in Fig. 12(a) and (b). As confirmed by our measurements (Fig. 12(c)), these miniaturized waveguide filter exhibit 5% and 10.4% bandwidth, respectively, for $W$=5 and $W$=7mm, with an average return loss (-$S_{11}$) of 15dB in the passband and an insertion loss below 0.7 dB (Fig. 12(d)).

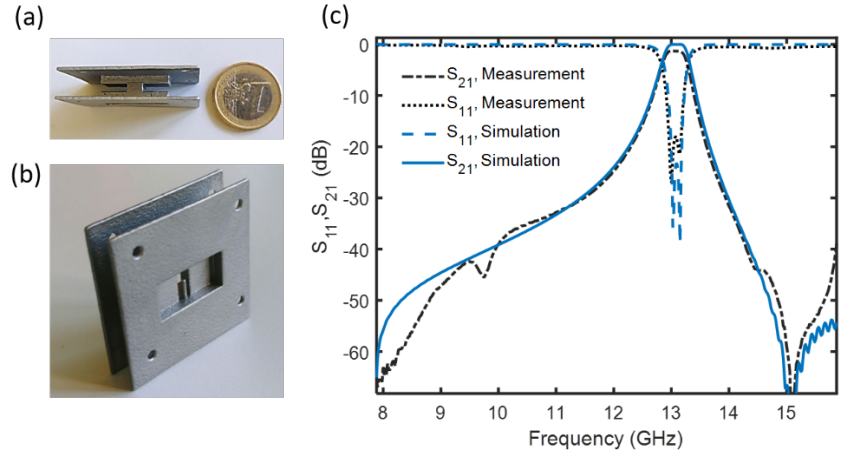

Fig. 11. (a) Top and (b) 3D view of the compact WR75 filter having two pins, (c) the measured and simulation results ($S_{21}$ and $S_{11}$).

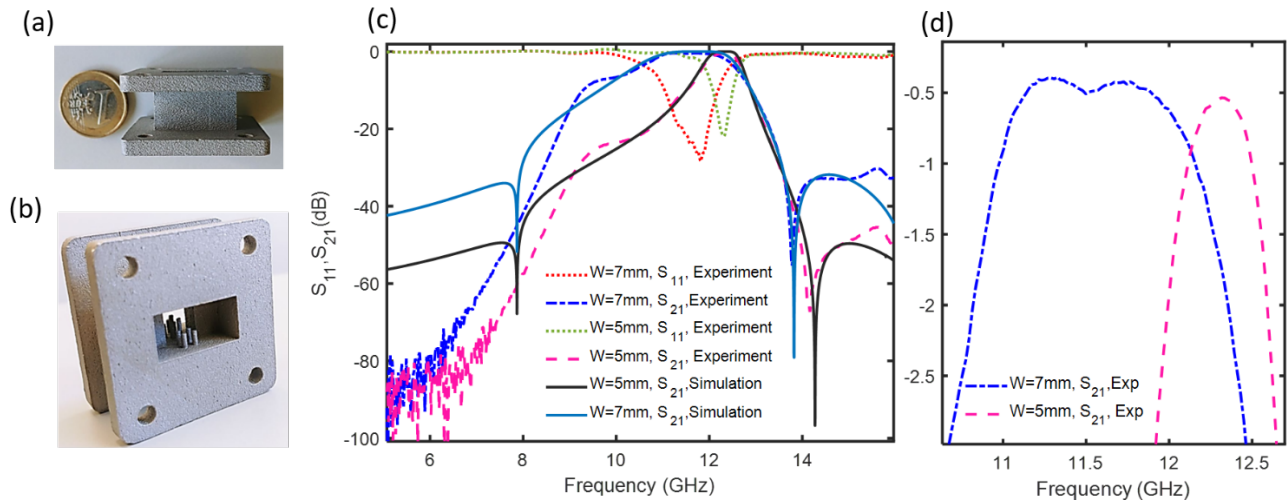

Fig. 12. (a) Top and (b) 3D view of 2$^{nd}$ order CPPW filter with thicker walls, including meta-ports ($h_p$=3.5mm). (c) The transmission spectra of the 2$^{nd}$ order filters, for $W$=5 and 7mm, extracted from simulation and experiments. (d) The measured $S_{21}$, which shows an insertion loss of 0.5dB.

Moving to higher-order bandpass filters with sharper roll-off, we consider a CPPW with an even higher number of pins and construct a practical preselect WR75 Ku-band filter with meta-ports, shown in Fig. 12(a). This filter can be used in broadband satellite communications, for example, for 5G or internet-over-satellite. Samples with different widths are fabricated based on SLM (AlSi10Mg aluminum alloy), and shown in Fig. 13 (a). The results of simulations and experimental measurements of three fabricated filters with various widths of *W*=4, 5, and 7.4mm are shown in Fig. 13 (b). By increasing *W* from 4mm to 7.4mm, the bandwidth enlarges from 4.5% to 17.5% for devices of similar footprints (0.76λ×1.28λ, where λ=25mm).

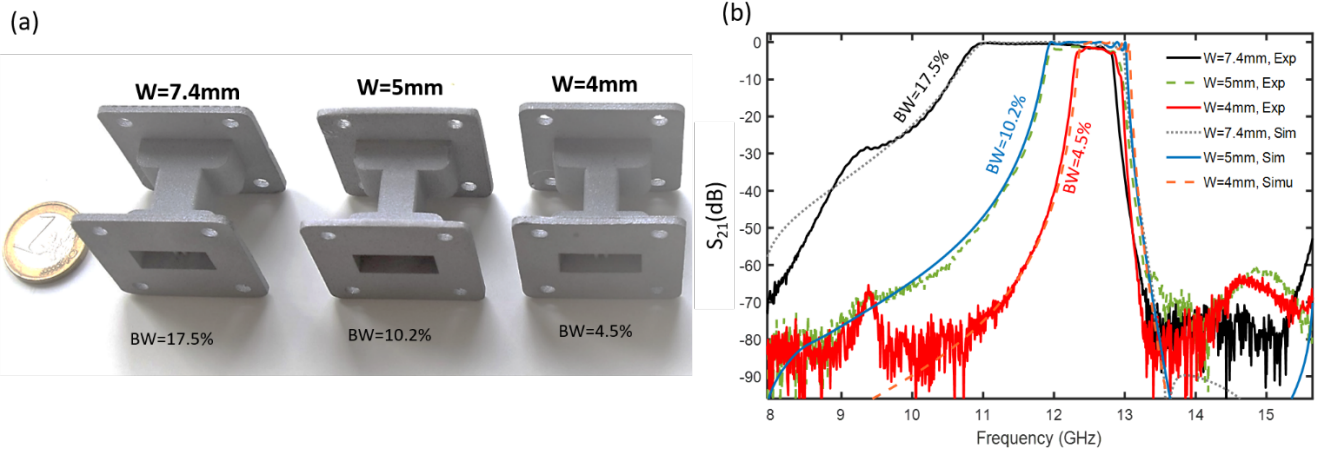

Fig. 13. (a) Fabricated 6$^{th}$ order CPPW filters with different widths *W*=3, 5, and 7.4mm, comprising metamaterial ports and improved matching. (b) The transmission spectra of the CPPW filters ($S_{21}$) extracted from simulation and experiments, indicating the fractional bandwidth (BW) in percent.

Next, we investigate the effect of geometrical scaling by shrinking down the CCPW filter by 20% in the longitudinal direction (Fig. 14(a)). It implies multiplying all lengths along *y* by 0.8, and the cross-sections of the pins become oval. The total length of the scaled component is accordingly reduced to 26mm, and its weight decreases from 26gr to 21gr, while the position of the passband remains unchanged (Fig. 14(b)), with untouched insertion loss (<0.7dB), rejection level (>60dB), and return loss (~15dB). This experiment indicates that the design may be open to further size optimization as long as it remains within reach of the manufacturing method. We also fabricated a silver-plated filter to reduce the insertion loss level, shown in the inset of Fig. 15. As shown in the figure, this silver-plated Ku band waveguide filter exhibits reduced insertion loss down to 0.25dB. Such competitive RF specifications establish CPPWs as high-performance waveguide filters yet with sizes at least one order of magnitude smaller than conventional waveguide filters in this frequency range, which is particularly remarkable.

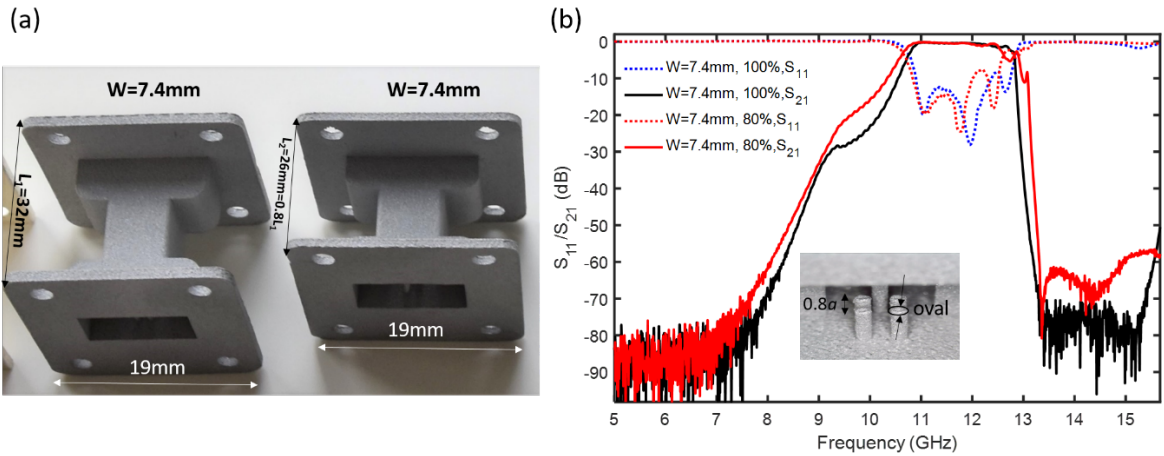

Fig. 14. (a) The manufactured standard CPPW filter and its down-scaled version (80% along y-direction), considering *W*=7.4mm. (b) The scattering spectra ($S_{21}$ and $S_{11}$) for the CPPW filter and its scaled-down version

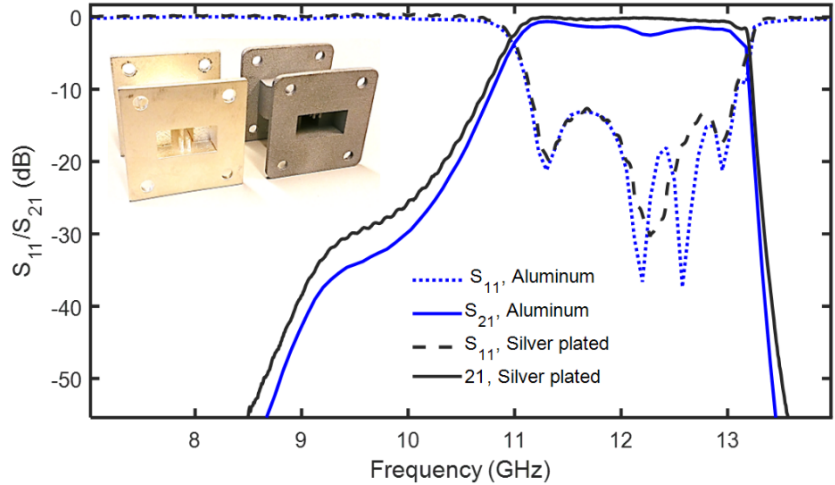

Fig. 15. The fabricated aluminum CPPW filters, with and without silver-plating, and their scattering spectra, highlighting the reduction of insertion losses after silver plating.

### III. CONCLUSION

We demonstrated an application of locally resonant metamaterials for creating metamaterial waveguide filters (Meta-filters), compatible with standard waveguide interfaces and with significantly reduced size and weight compared to current solutions. To illustrate the capabilities of the proposed meta-filter, we introduced the exemplary model of composite pin-pipe waveguides (CCPWs). We showed that their passband is strongly related to the cut-off frequency of the host waveguide, and that the concept is also compatible with periodic and random resonator arrangements, as well as with various types of ports. By proposing subwavelength metamaterial ports (Meta-ports), we improved the matching between such a CPPW filter and standard rectangular waveguides. Finally, we fabricated low- and high-order CPPW bandpass filters with various bandwidths, ranged from 4.5% to 17.5%, in a small and fixed footprint. We also manufactured a scaled version of a bandpass filter with 80% size reduction, which keeps its RF specifications, and confirmed silver-plating as a viable option to optimize insertion losses. Our experimental measurements on 3D-printed Ku band filters exhibit competitive insertion loss (down to 0.25dB), rejection levels (>60dB) and return loss (~15dB). This study demonstrates the relevance of the CPPW concept to realize efficient devices with unprecedented small sizes, small volume and lightweight, while maintaining the compatibility with standard waveguide ports, and without sacrificing the RF specifications when compared to traditional designs.

### Appendix I. Guiding mechanism of CPPWs in the evanescent regime

We propose a distributed circuit model to illustrate the effect of the various coupling mechanisms that take part in the creation of a guided mode. Contrary to microstrip metamaterial filters [8,28,29], and planar interdigital filters, which can often be modeled using a lumped circuit model, here we must employ a distributed elements circuit to model both finite and infinite CPPWs. Such a model is useful for explaining the guiding mechanism at play and determining the effect of pipe on realizing custom bandwidths. In a general model, a CPPW is composed on a cluster of coupled resonant pins and an evanescent pipe (Fig. S1(a)). As shown in Fig. S1(b), we assume resonant pins as shunt LC tanks, directly electrically and magnetically coupled (Fig.S1(b)) while higher-order modes can exist. These couplings are represented by mutual capacitance ($C_{ij}$) and a mutual inductance ($L_{ij}$) parameters ($m_{ij} = j\omega C_{ij} + \frac{1}{j\omega L_{ij}}$) in this microscopic model, while their macroscopic effect is to contribute to the creation of the HBG.

Besides, we need to use a distributed model for a short length of the evanescent pipe, with length *l*. The lumped LC circuit model, such as what is used for metamaterial planar devices, is not applicable here. For modeling an evanescent waveguide, we can use the inductive π circuit model, shown in Fig. S1(c) [2]. The elements of the distributed model can be calculated based on a real propagating constant k and an imaginary impedance $Z_0=jX_0$. By loading coupled deep subwavelength resonators inside the evanescent pipe, we obtain Fig.S1(d), where both the effects of direct EM coupling and evanescent host are considered, but each one alone is not sufficient to explain the guiding mechanism. The coupling of the energy between different meta-atoms can equivalently be pictured by defining electric and magnetic coupling coefficients κ$_E$ and κ$_M$, and an evanescent coupling coefficient κ$_{evan}$, which can be written as κ$_{evan}$=1/cosh(*kl*), where $k = \frac{2\pi f}{c} \cdot \sqrt{(f_c/f)^2 - 1}$ is the propagation constant of the evanescent waveguide, and *l* is distances between the pins (*l=a-2r*). The primary parameters that can be used for adjusting the coupling coefficients include the distances separating the pins (*l=a*) and the pipe width *W*, while the latter changes the value of κ$_{total}$=κ$_E$ +κ$_M$ + κ$_{evan}$ significantly. Although the κ$_E$ and κ$_M$ can be computed numerically based on equations reported in [30], here, we focus on κ$_{evan}$, since it dominates the coupling of pins when separated by deep subwavelength distances and strongly coupled. As shown in Fig.S2, reducing the separating distance results in a higher κ$_{evan}$. Nevertheless, this effect is not as effective as increasing the width of the pipe. Thus, varying *W* is the most straightforward solution to construct both narrow and wideband CPPWs, where the pipe is below the cut-off.

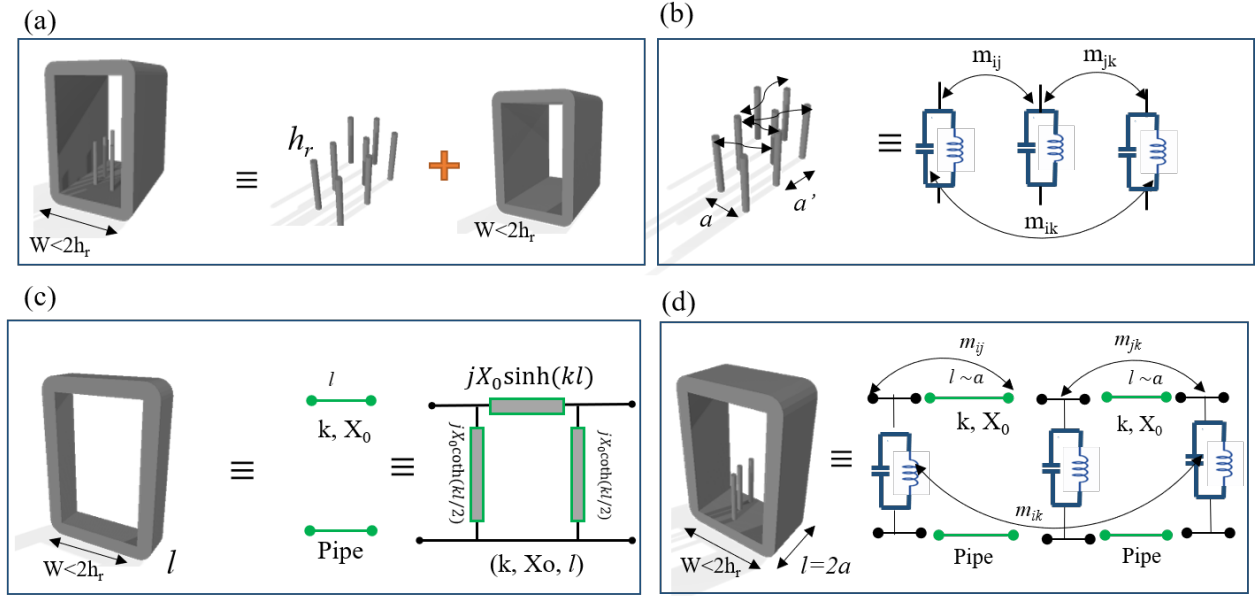

FIG. S1. (a) Decomposition of a CPPW in evanescent regime into direct-coupled pins and an evanescent pipe. (b) The lumped circuit model of the coupled resonant pins, where $m_{ij} = j\omega C_{ij} + \frac{1}{j\omega L_{ij}}$. (c) The distributed element model of an evanescent pipe. (d) The general distributed model of CPPW.

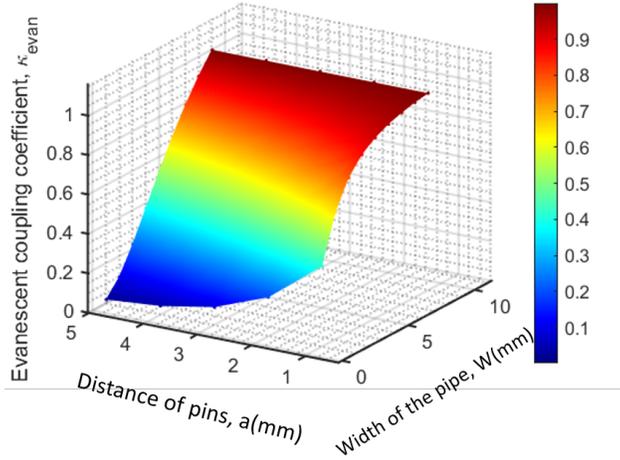

FIG. S2. The variation of evanescent coupling versus pin separation distance $a$ and width of the pipe $W$.

### Appendix II. Differences between CPPW and combline filters

CPPWs must not be confused with another type of waveguide filters involving circular rods, known as combline filters. CPPWs are based on a totally different guiding mechanism, involving strong spatial dispersion (multiple scattering), whereas combline filters do not involve spatial dispersion, and as a matter of fact are always described with local circuit models, leading to markedly different features and size constraints. We explain here in details the difference between CPPWs and combline filters.

Combline filters are designed using parallel quarter-wavelength coaxial cavities, where the impedance, quality factor, and resonance frequency of the cavities are calculated based on coupled TEM modes, which are absent in CPPWs. Unlike for CPPWs, the operating frequency in combline filters strongly depends on the cross-section of internal rods or ridge, the size of the surrounding box, and the air gap between the inner rods and the top plate. The rods diameter ($2r$) is usually selected to obtain a highest quality factor; thus, it is chosen in a range of $0.2W$ to $0.5W$, where $W$ is the distance between the ground plates. The numerical results from TEM theory for larger values of $W$ ($W/\lambda > 0.08$) show significant deviation from experiment [31]. This deviation is initially because of the limitations of TEM mode theory for modeling the structures with wider widths, especially in higher frequency. Thus, the combline filter model is completely inadequate and not applicable in the case of deep-subwavelength structures such as the CPPWs proposed in this paper ($2r<<W$, $2r<<\lambda$).

Contrary to combline filters, CPPWs are composed of deep subwavelength coupled pins, having a cross-section much smaller than the waveguide width ($2r<<W$). The interaction of host modes and local resonances is exploited around the resonance frequency of the pins, for creating customizable pass and rejection bands. Since the width of the pipe is in the range of $0.08\lambda < W < \lambda/2$ and the cross-sections of the pins are very small, the local combline filter theory is not adequate. For example, it is not possible to model the structure using a local model, with self and mutual capacitances. The behavior of a CPPW can be captured using more advanced models, or full-wave simulations that will correctly consider the spatial dispersion and multiple scattering of the resonant elements. This makes CPPWs a markedly different paradigm.

The main feature of CPPW filters, demonstrated in section II, is that the induced HBG (rejection band) is independent of the arrangement and distances of elements, and it is also decoupled from the size and shape of the hollow pipe or the surrounding box. However, the passband bandwidth can be altered by changing the pipe width, which is a striking feature. All these properties are absent in combline filters, for which the positions of the pass and stopbands are sensitive to the size of the surrounding cavity and air gap between the rod and the top plate. For a brief example of this, we compute the band diagrams of and infinite periodic CPPW, with a unit cell shown in Fig. S3(a) and (b), for various values of $h_g$. We see that the HBG is not sensitive to $h_g$ (Fig.S3(b)). We compare this behavior to a basic design of a combline filter unit-cell (Fig. S1 (c)) and (d), using a rod with the same height ($h_r$=5mm). The extracted band diagram for various values of $h_g$ exhibit a completely different behavior when compared to CPPWs. Figure S3(d), shows that the band edge of the guided modes varies significantly by altering the values of $h_g$. Thus, the operating frequency is susceptible to the air gap. Besides, Fig.S3(e) and (f) show that for a 6$^{th}$ order coaxial combline filter (total length 6$W$); the operating frequency is shifting when altering the width of the hollow metallic box, forcing the total system to scale with the wavelength of operation, unlike CPPWs (total length 6$a$). These differences are due to the different guiding mechanisms at play in the two distinct types of filters.

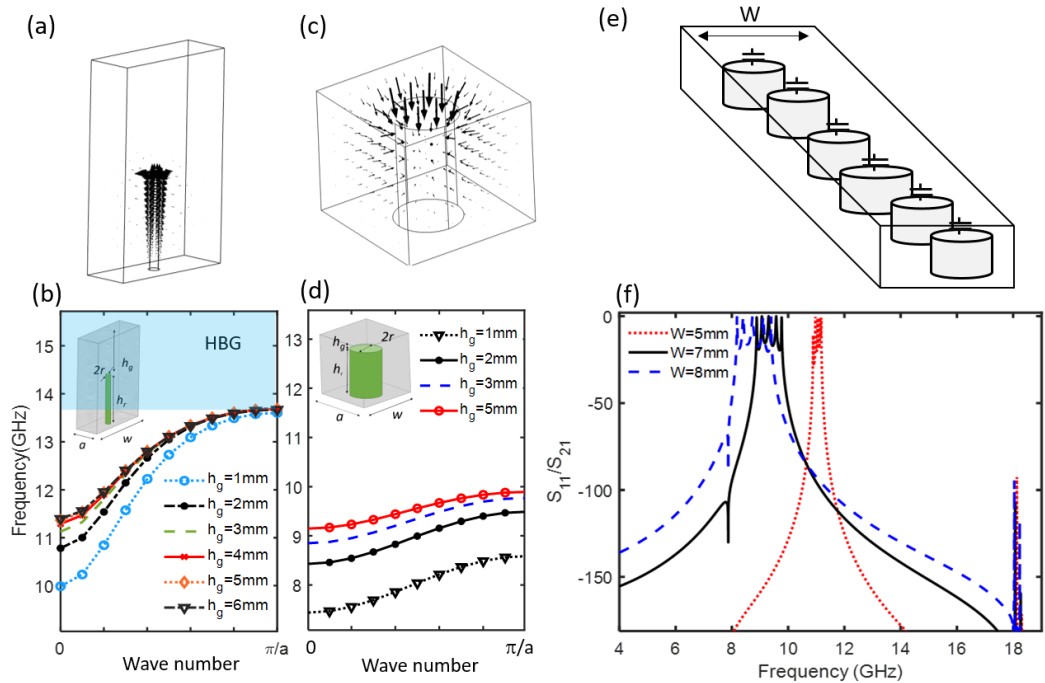

FIG. S3 (a) A unit-cell of a CPPW, composed of a thin pin ($r<<W$ and $r<<a$), $W$=8mm, $h_r$=5mm, (b) Band diagram of the CPPW for various values of $h_g$, which shows the HBG for all cases and a converged sub-$\lambda$ mode for $h_g$>3mm. (c) A coaxial cavity, having $a=W$=8mm, $h_r$=5mm, and $2r=W/2$. (d) Sensitivity of the guided mode dispersion to the top gap size in an infinite array of coaxial unit-cells. (e) A 6$^{th}$ order combline filter with $h_r$=5mm, and $2r=W/2$. (f) The transmission spectra of the combline filter with a total length of 6$W$ for various values of $W$. Very different from CPPWs, the frequency band of combline filters shifts significantly when $h_g$ or $W$ is varied, evidencing a very different guiding mechanism.

## References

[1] V. E. Boria and B. Gimeno, *Waveguide Filters for Satellites*, IEEE Microw. Mag. **8**, 60 (2007).
[2] G. L. Matthaei, B. Schiffman, E. Cristal, and L. Robinson, *Microwave Filters and Coupling Structures*, (1963).
[3] C. Kudsia, R. Cameron, and W. Tang, *Innovations in Microwave Filters and Multiplexing Networks for Communications Satellite Systems*, **40**, (1992).
[4] R. Levy and S. B. Cohn, *A History of Microwave Filter Research, Design, and Development*, IEEE Trans. Microw. Theory Tech. **32**, 1055 (1984).
[5] G. F. Craven and C. K. Mok, *The Design of Evanescent Mode Waveguide Bandpass Filters for a Prescribed Insertion Loss Characteristic*, IEEE Trans. Microw. Theory Tech. **19**, 295 (1971).

[6] R. Engheta, Nader, Ziolkowski, *Metamaterials: Physics and Engineering Explorations* (Wiley, 2006).
[7] R. Marques, F. Martin, and M. Sorolia, *Metamaterials with Negative Parameters* (Wiley, New Jersey, 2008).
[8] M. Gil, J. Bonache, and F. Martín, *Metamaterial Filters: A Review*, Metamaterials **2**, 186 (2008).
[9] F. Martin, *Artificial Transmission Lines for RF and Microwaves Applications* (Wiley, New Jersey, 2015).
[10] C. Wenshan and V. Shalaev, *Optical Metamaterials: Fundamentals and Applications* (Springer, 2009).
[11] F. Lemoult, N. Kaina, M. Fink, and G. Lerosey, *Wave Propagation Control at the Deep Subwavelength Scale in Metamaterials*, Nat. Phys. **9**, 55 (2013).
[12] N. Kaina, F. Lemoult, M. Fink, and G. Lerosey, *Ultra Small Mode Volume Defect Cavities in Spatially Ordered and Disordered Metamaterials*, Appl. Phys. Lett. **102**, (2013).
[13] N. Kaina, A. Causier, Y. Bourlier, M. Fink, T. Berthelot, and G. Lerosey, *Slow Waves in Locally Resonant Metamaterials Line Defect Waveguides*, Sci. Rep. **7**, 15105 (2017).
[14] Z. Liu, X. Zhang, Y. Mao, and Y. Y. Zhu, *Locally Resonant Sonic Materials*, Science (80-. ). **289**, 1734 (2000).
[15] B. Orazbayev and R. Fleury, *Quantitative Robustness Analysis of Topological Edge Modes in C6 and Valley-Hall Metamaterial Waveguides*, Nanophotonics **8**, 1433 (2019).
[16] M. K. Moghaddam and R. Fleury, *A Subwavelength Microwave Bandpass Filter Based on a Chiral Waveguide*, 14th Eur. Conf. Antennas Propag. **1**, (2020).
[17] M. K. Moghaddam and R. Fleury, *Slow Light Engineering in Resonant Photonic Crystal Line-Defect Waveguides*, Opt. Express **27**, 26229 (2019).
[18] M. L. Cowan, J. H. Page, and P. Sheng, *Ultrasonic Wave Transport in a System of Disordered Resonant Scatterers: Propagating Resonant Modes and Hybridization Gaps*, Phys. Rev. B - Condens. Matter Mater. Phys. **84**, 1 (2011).
[19] A. Rahimi-iman and P. Lasers, *Springer Series in Optical Sciences 229 Polariton Physics* (n.d.).
[20] J. B. Pendry, A. J. Holden, D. J. Robbins, and W. J. Stewart, *Low Frequency Plasmons in Thin-Wire Structures*, J. Phys. Condens. Matter **10**, 4785 (1998).
[21] D. R. Smith, J. B. Pendry, M. C. K. Wiltshire, and X. Zhang, *Metamaterials and Negative Refractive Index*, Science (80-. ). **305**, 788 (2004).
[22] David M. Pozar, *Microwave Engineering*, 4th ed. (New Jersey, 1998).
[23] https://www.etlsystems.com/.
[24] P. Vallerotonda, L. Pelliccia, C. Tomassoni, F. Cacciamani, R. Sorrentino, J. Galdeano, and C. Ernst, *Compact Waveguide Bandpass Filters for Broadband Space Applications in c and Ku-Bands*, Proc. Eur. Microw. Conf. Cent. Eur. EuMCE 2019 116 (2019).
[25] P. S. Kildal, *Artificially Soft and Hard Surfaces in Electromagnetics*, IEEE Trans. Antennas Propag. **38**, 1537 (1990).
[26] A. Polemi, S. MacI, and P. S. Kildal, *Dispersion Characteristics of a Metamaterial-Based Parallel-Plate Ridge Gap Waveguide Realized by Bed of Nails*, IEEE Trans. Antennas Propag. **59**, 904 (2011).
[27] O. A. Peverini, G. Addamo, M. Lumia, G. Virone, F. Calignano, M. Lorusso, and D. Manfredi, *Additive Manufacturing of Ku/K-Band Waveguide Filters: A Comparative Analysis among Selective-Laser Melting and Stereolithography*, IET Microwaves, Antennas Propag. **11**, 1 (2017).
[28] J. D. Baena, J. Bonache, F. Martín, R. M. Sillero, F. Falcone, T. Lopetegi, M. A. G. Laso, J. García-García, I. Gil, M. F. Portillo, and M. Sorolla, *Equivalent-Circuit Models for Split-Ring Resonators and Complementary Split-Ring Resonators Coupled to Planar Transmission Lines*, IEEE Trans. Microw. Theory Tech. **53**, 1451 (2005).
[29] J. García-García, F. Martín, F. Falcone, J. Bonache, J. D. Baena, I. Gil, E. Amat, T. Lopetegi, M. A. G. Laso, J. A. M. Iturmendi, M. Sorolla, and R. Marqués, *Microwave Filters with Improved Stopband Based on Sub-Wavelength Resonators*, IEEE Trans. Microw. Theory Tech. **53**, 1997 (2005).
[30] K. Awai and Y. Zhang, *Coupling Coefficient of Resonators - An Intuitive Way of Its Understanding*, Electron. Commun. Japan, Part II Electron. (English Transl. Denshi Tsushin Gakkai Ronbunshi) **90**, 11 (2007).
[31] R. Levy, H. Yao, and K. A. Zaki, *Transitional Combline / Evanescent-Mode Microwave Filters*, **45**, 2094 (1997).